\newif\ifonecol
\onecolfalse

\documentclass[journal,10pt]{IEEEtran}
\usepackage{epsfig}
\usepackage{epstopdf}
\usepackage{amssymb}
\usepackage{amsmath}
\usepackage{cite}
\usepackage{cases}
\usepackage{algpseudocode}
\usepackage{algorithm}
\usepackage{dsfont}
\usepackage{multirow}
\usepackage{amsmath}
\usepackage{enumerate}
\usepackage{color}

\newcommand{\argmin}{\operatornamewithlimits{argmin}}

\begin{document}
\title{{Knowledge Distillation-aided End-to-End Learning for Linear Precoding in Multiuser MIMO Downlink Systems with Finite-Rate Feedback}}

\author{Kyeongbo Kong, Woo-Jin Song \emph{Member, {IEEE}}, and Moonsik Min \emph{Member, {IEEE}\vspace{-6mm}}
\thanks{Kyeongbo Kong and Woo-Jin Song are with the Department of Electrical Engineering, Pohang University of Science and Technology, Pohang 37673, South Korea (e-mail: kkb4723@postech.ac.kr, wjsong@postech.ac.kr), and Moonsik Min is with the School of Electronics Engineering, Kyungpook National University, Daegu 41566, South Korea (e-mail: msmin@knu.ac.kr).}
}

\markboth{}{}

\maketitle

\begin{abstract}
We propose a deep learning-based channel estimation, quantization, feedback, and precoding method for downlink multiuser multiple-input and multiple-output systems.
In the proposed system, channel estimation and quantization for limited feedback are handled by a receiver deep neural network (DNN).
Precoder selection is handled by a transmitter DNN.
To emulate the traditional channel quantization, a binarization layer is adopted at each receiver DNN, and the binarization layer is also used to enable end-to-end learning.
However, this can lead to inaccurate gradients, which can trap the receiver DNNs at a poor local minimum during training. 
To address this, we consider knowledge distillation, in which the existing DNNs are jointly trained with an auxiliary transmitter DNN.
The use of an auxiliary DNN as a teacher network allows the receiver DNNs to additionally exploit lossless gradients, which is useful in avoiding a poor local minimum.
For the same number of feedback bits, our DNN-based precoding scheme can achieve a higher downlink rate compared to conventional linear precoding with codebook-based limited feedback. 
\end{abstract}


\IEEEpeerreviewmaketitle

\vspace{-2mm}
\section{Introduction}
In multiuser multiple-input and multiple-output (MU-MIMO) downlink systems with frequency-division duplexing, the transmitter cannot track the downlink channel directly.
Thus, finite-rate feedback (or limited feedback) is widely used to achieve partial channel state information (CSI) at the transmitter (CSIT) \cite{Jindal2006, Ravindran2008, Min2016}.
The performance of such systems is significantly affected by the CSIT accuracy, which is determined by the number of feedback bits. 
Therefore, state-of-the-art commercial mobile devices for cellular networks do not actively use MU-MIMO downlink applications, despite the 4G long-term evolution (LTE)-advanced standards defining up to four layers, and the recently released 5G new radio (NR) standards defining up to 12 layers for these applications \cite{Mondal2019}. 
This is mainly because the achievable rate is not sufficiently high to compensate for the cost of the uplink resources that convey the feedback bits.

Recently, deep learning (DL)-based methods have been proposed to improve the performance of limited feedback for MIMO systems.
Researchers designed a CSI sensing and recovery network (CsiNet) using a convolutional neural network to improve the accuracy of the CSIT \cite{Wen2018}. 
A recurrent neural network (RNN) which captures the time correlation to enhance the channel recovery module was developed \cite{Wang2019}. 
Recurrent compression and uncompression modules to improve an RNN structure were studied \cite{Lu2019}.
Using CSI compression with quantization through a bit-level optimized neural network (NN), researchers improved the performance of the CSI compression \cite{lu2019bit}.
A joint optimization of the CSI compression, quantization, and recovery was considered to improve CSI accuracy \cite{liu2020efficient}. 
A compressive sensing NN framework was proposed to effectively compress and quantize the CSI \cite{guo2020convolutional} . 
A distributed CSI compression and feedback framework that benefits from the correlations between the nearby users was studied \cite{guo2020dl, mashhadi2020distributed}. 
Applications of the DL-based CSI quantization methods to more complicated multi-cell scenarios were considered \cite{guo2020deep}.
However, these studies have assumed a perfect CSI at the receiver (i.e., they did not consider channel estimation).



To enable a joint optimization of the channel estimation and quantization, we employ a DL-based limited feedback and beamforming model proposed in \cite{Jang2020}, which was originally intended for a single-user MIMO channel.
First, we extend the system to the full-stream MU-MIMO downlink systems with limited feedback.
In the proposed system, individual deep neural networks (DNNs) are implemented for the users and the base station (BS).
Similar to the previous study \cite{Jang2020}, we adopt a binarization layer at each receiver DNN to enable end-to-end learing. 
Parallel to our work, this framework was extended to a multiuser massive MIMO scenario by another research group that designed joint channel estimation, quantization, power allocation, and beamforming selection in mmWave massive MIMO channels, assuming a sufficient number of feedback bits for a CSIT \cite{Sohrabi2020}.
The corresponding results have potential for the further evolution of 5G communications.
However, as the 5G standard does not exclude the previous LTE/LTE-advanced (LTE-A) standards in attempt to retain compatibility, studies dealing with the activation of MU-MIMO systems for operational commercial BSs and mobile devices are also important. 
The state-of-the-art BSs and mobile devices (such as smartphones in the market) are simultaneously activating fewer than eignt antennas (consequently, the available number of layers is fewer than $8$) for downlink MU-MIMO, with a relatively low number of feedback bits for the CSIT.
This study assumes such a practical configuration of the number of antennas and feedback bits, and proposes an efficient joint channel estimation, feedback, and precoding scheme using DNNs that outperforms the conventional limited-feedback-based precoding systems.
Moreover, we focus on improving the performance of an end-to-end training that is based on the use of a binarization layer.

Specifically, the binarization operation in the corresponding layer can cause a vanishing gradient problem in backpropagation during the end-to-end learning.
To address this, recent studies on the finite-rate NN-based CSI feedback in  MIMO systems \cite{guo2020convolutional, liu2020efficient, Sohrabi2020, guo2020deep, Jang2020, guo2020dl, lu2019bit} employed pseudo gradients of a straight-through estimator (STE).
However, the pseudo-gradients may not be in the right direction for updating the parameters, thereby trapping the DNNs at a poor local minimum \cite{zhuang2018, zhuang2019}.
To overcome this problem, we propose a joint training method with knowledge distillation (KD) in which the receiver DNNs are effectively trained by using additional ``lossless gradients'' with the aid of an auxiliary teacher network.
Subsequently, an end-to-end learning is jointly performed to determine the precoding matrices that maximize the downlink sum rate.
The proposed data-driven solution for the precoding matrices outperforms the conventional linear precoding methods.
In addition, 
our training method with KD can be applied to the studies that use STEs (such as \cite{guo2020convolutional, liu2020efficient, Sohrabi2020, guo2020deep, Jang2020, guo2020dl, lu2019bit}), to improve the performance of the training process.

\section{System Model}
\subsection{Channel model and performance metric}
This study considers linear precoding methods in MU-MIMO downlink channels based on limited feedback.
In particular, we consider a MIMO broadcast channel in which $K$ users communicate with a single BS.
The BS is equipped with $M$ transmit antennas, and each user has $N$ receive antennas.
We assume $M$ to be a multiple of $N$ where $M>N$.
The MIMO channels between the BS and users (users from the set $\{1,\ldots,K\}$) are modeled as independent channel matrices
$\mathbf{H}_k\in \mathds{C}^{M\times N}$, for $k=1,\cdots, K$, where the entries of each matrix are independent and identically distributed (i.i.d.) complex Gaussian random variables, and each entry has a zero mean and unit variance.
The received signal at user $k$ is given by $\mathbf{y}_k = \mathbf{H}_{k}^H \mathbf{x} + \mathbf{n}_k$, for 
$k=1,\cdots, K$, where $\mathbf{x}$ is the transmit vector and $\mathbf{n}_k$ is a complex Gaussian noise vector with independently distributed entries; each entry has a zero mean and unit variance.
The transmit vector is expressed as  $\mathbf{x}=\sum_{l=1}^{K}\mathbf{V}_l\mathbf{s}_l$, where $\mathbf{V}_l \in \mathds{C}^{M\times N}$ is the precoding matrix; we assume $\text{tr}(\mathbf{V}_l^H\mathbf{V}_l)=N$ to realize an equal power allocation for users.
The information symbol vector $\mathbf{s}_l \in \mathds{C}^{N\times 1}$ consists of $N$ independent data symbols for user $l$
so that each user is fully served with $N$ degrees of spatial multiplexing.
In addition, we do not consider specific user selection, and assume that $K\leq M/N$.
We impose the power constraint $P$ at the BS, and the transmit power allocation for the data streams is $\mathbb{E}[\mathbf{s}_l\mathbf{s}_l^H]= \frac{P}{KN}\mathbf{I}_N$ to provide an equal power allocation to users.
The equal power allocation is considered in this study for a fair comparison between the baseline and proposed schemes, and to simply realize fair resource allocation among users. 
Therefore, the achievable rate of user $k$ is given by 
\begin{align}
    \label{eq5}
    R_{k}
    &\triangleq\mathbb{E}[\log_{2}{\vert \mathbf{I}_N + \frac{P}{M}\sum_{{l=1}}^{K}\mathbf{H}_k^H{\mathbf{V}}_l{\mathbf{V}}_l^H\mathbf{H}_k\vert}] \nonumber \\
    &-\mathbb{E}[\log_{2}{\vert \mathbf{I}_N + \frac{P}{M}\sum_{\substack{l=1, l\neq k}}^{K}\mathbf{H}_k^H{\mathbf{V}}_l{\mathbf{V}}_l^H\mathbf{H}_k\vert}].
\end{align}




\subsection{Baseline scheme: limited-feedback-based linear precoding}




Before a CSI feedback, each user first estimates the channel using a pilot \cite{Jang2020}. 
Let $\mathbf{p}_l\in\mathbb{C}^{M\times 1}$, for $l=1,\cdots, L$, be the normalized pilot sequences. 
Subsequently, the received signal $\mathbf{y}_{l,k}^{\text{train}}=\sqrt{P_{\text{train}}}\cdot\mathbf{H}_{k}^H\mathbf{p}_l+\mathbf{n}_k$, which corresponds to the pilot transmission for $l=1,\cdots, L$ is used to estimate $\mathbf{H}_{k}$, where $P_{\text{train}}$ is the transmit power for the pilot. 
The estimated channel $\bar{\mathbf{H}}_{k}$ is obtained using the minimum mean square error estimation, where the $i$-th component of $\mathbf{p}_l$ is given by $\frac{1}{\sqrt{M}}e^{j2\pi(i-1)(l-1)/L}$ \cite{Biguesh2006}.

Let $\bar{\mathbf{H}}_k = \widetilde{\mathbf{H}}_k\mathbf{\Sigma}_k^{\frac{1}{2}}\mathbf{U}_k^H$ represent the compact SVD of the estimated channel matrix.
Since quantizing the $\bar{\mathbf{H}}_k$ itself is inefficient (it requires a large number of feedback bits), the conventional limited feedback model quantizes and feeds back the unitary matrix $\widetilde{\mathbf{H}}_k\in\mathbb{C}^{M\times N}$ that contains the direction information of the channel \cite{Ravindran2008}.
User $k$ quantizes $\widetilde{\mathbf{H}}_k$ using the codebook $\mathcal{C}_k = \left\{\mathbf{A}_{k,1}, \ldots, \mathbf{A}_{k,2^B}\right\}$, which is fixed beforehand for each user and is known to the transmitter. Each codeword $\mathbf{A}_{k,j}$ is given by a semi-unitary matrix in $\mathds{C}^{M\times N}$ such that $\mathbf{A}_{k,j}^H\mathbf{A}_{k,j}=\mathbf{I}_{N}$, and is different from all other codewords.
With $\mathcal{J}=\{1,\ldots, 2^B\}$, the quantization process is given by 
\begin{align}
    \label{eq2}
    {q}_k = \argmin_{j\in\mathcal{J}}
    d\left(\mathbf{A}_{k,j}, \widetilde{\mathbf{H}}_k\right),
\end{align} where $d(\cdot, \cdot)$ is the distance measure.
Each user feeds back index $q_k$ to the transmitter, and the transmitter obtains a quantized channel matrix $\widehat{\mathbf{H}}_{k} = \mathbf{A}_{k, q_k}$ from codebook $\mathcal{C}_k$.
The transmitter constructs precoding matrices based on quantized channels fed back from the users.

\begin{figure*}
    \centering
    {\epsfig{file=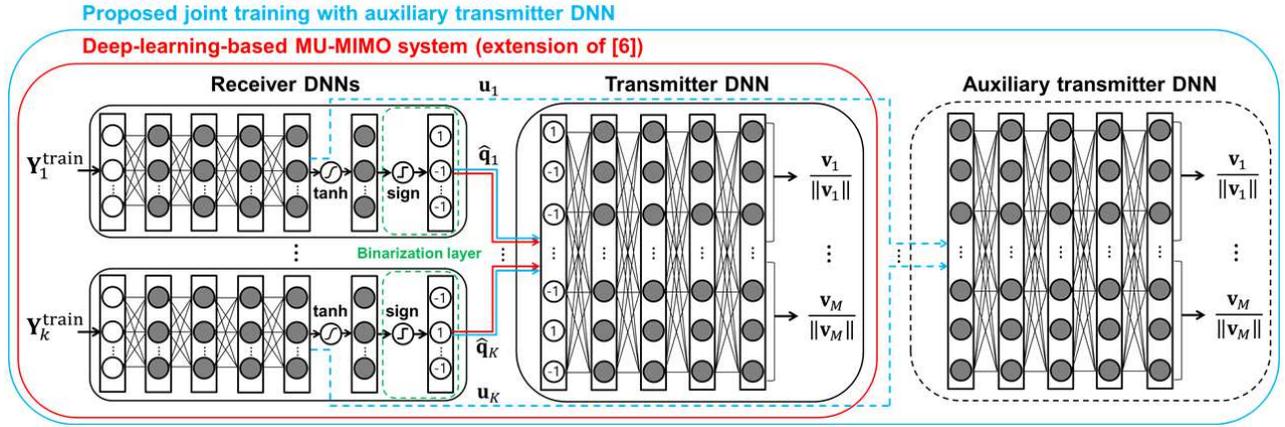,width=17cm}} 
    \caption{Schematic of the proposed DL-based MU-MIMO system with a joint training method. In the proposed system, receiver DNNs estimate the feedback information for each input channel and the estimated information is used as input to the transmitter DNN (solid red line). The transmitter DNN estimates the precoding matrices, where the columns of each precoding matrix are normalized. In the training phase, receiver DNNs are connected with both the transmitter DNN (solid blue line) and auxiliary transmitter DNN (dotted blue line).
    Then, a joint training is performed to utilize a lossless gradient from the auxiliary DNN to the receiver DNNs, as well as the lossy gradient (due to binarization layer) from the transmitter DNN to the receiver DNNs.\vspace{-3mm}}
    \label{fig8}
\end{figure*}

In this study, we consider normal \cite{Ravindran2008} and regularized block diagonalizations \cite{Stankovic2008} as baseline linear precoding schemes, since they have a high efficiency among the existing linear precoding schemes with limited feedback in terms of the achievable rate per feedback bit.
If the transmitter has a perfect CSI, an iterative method called weighted minimum mean squared error (WMMSE) precoding provides a local optimal solution for the linear precoding design \cite{Christensen2008}.
Similarly, other iterative methods based on the regularized BD (RBD) to achieve a near-optimal performance have been studied \cite{Stankovic2008}.
However, to realize such a near-optimal linear precoding based on iterations, the transmitter must have an accurate information of the entire channel matrix $\mathbf{H}$ rather than the CDI $\mathbf{\widetilde{H}}_k$, i.e, an additional feedback is required to send the quantized information of $\mathbf{\Sigma}_k$.
Moreover, the sum rate enhancement largely relies on the dynamic power allocation among users for successive iterations.
Thus, these iterative approaches are not suitable for limited-feedback-based precoding systems that operates based on CDI feedback.
Therefore, we do not consider iterative methods such as the WMMSE or iterative RBD in this study.

To obtain the RBD precoder (Section II-B of \cite{Stankovic2008}), we first consider the following auxiliary matrix: $\mathbf{\Phi}_k=[\widehat{\mathbf{H}}_{1}\;\cdots \;\widehat{\mathbf{H}}_{{k-1}}\;\;\widehat{\mathbf{H}}_{{k+1}}\;\cdots\; \widehat{\mathbf{H}}_{K}]^H$.
Let the SVD of $\mathbf{\Phi}_k$ be $\mathbf{\Phi}_k=\mathbf{E}_{k,0}\mathbf{D}_{k,0}(\mathbf{G}_{k,0})^H$ such that $\mathbf{D}_{k,0}\in\mathbb{C}^{N(K-1)\times M}$ and $\mathbf{G}_{k,0}\in\mathbb{C}^{M\times M}$.
Further, let $\widetilde{\mathbf{G}}_{k,0}\in\mathbb{C}^{M\times (M-N(K-1))}$ be the matrix whose $i$-th column is equal to the $(N(K-1)+i)$-th column of $\mathbf{G}_{k,0}$ for $1\leq i\leq M-N(K-1)$, and let $\mathbf{F}_k^{\text{BD}} = \widehat{\mathbf{H}}_k^H\widetilde{\mathbf{G}}_{k,0}\in\mathbb{C}^{N\times (M-N(K-1))}$.
Expressing the compact SVD of $\mathbf{F}_k^{\text{BD}}$ as $\mathbf{F}_k^{\text{BD}}=\mathbf{E}_{k,1}\mathbf{D}_{k,1}(\mathbf{G}_{k,1})^H$ (such that $\mathbf{D}_{k,1}\in\mathbb{C}^{N\times N}$ and $\mathbf{G}_{k,1}\in\mathbb{C}^{(M-N(K-1))\times N}$), the BD precoding matrix is given by $\mathbf{V}_{k}=\widetilde{\mathbf{G}}_{k,0}\mathbf{G}_{k,1}$.
Furthermore, let $\mathbf{F}_{k,0}^{\text{RBD}}
    = \widehat{\mathbf{H}}_k^H\mathbf{G}_{k,0}(\mathbf{D}_{k,0}^H\mathbf{D}_{k,0}+\alpha_{\text{BD}}\cdot\mathbf{I})^{-\frac{1}{2}}\in\mathbb{C}^{N\times M}$, 
where $\alpha_{\text{BD}}$ is the regularization parameter for the BD.
Denoting the compact SVD of $\mathbf{F}_{k,0}^{\text{RBD}}$ as $\mathbf{F}_{k,0}^{\text{RBD}} =\mathbf{E}_{k,2}\mathbf{D}_{k,2}(\mathbf{G}_{k,2})^H$ (such that $\mathbf{D}_{k,2}\in\mathbb{C}^{N\times N}$ and $\mathbf{G}_{k,2}\in\mathbb{C}^{M\times N}$), we define
$\mathbf{F}_{k,1}^{\text{RBD}}
    \triangleq \mathbf{G}_{k,0}(\mathbf{D}_{k,0}^H\mathbf{D}_{k,0}+\alpha_{\text{BD}}\cdot\mathbf{I})^{-\frac{1}{2}}\mathbf{G}_{k,2}.$
On considering an equal power allocation to the users, the RBD precoding matrix for user $k$ is given by $\mathbf{V}_{k}=\sqrt{\frac{N}{\text{tr}\left(\mathbf{F}_{k,1}^{\text{RBD}}\right)}}\cdot\mathbf{F}_{k,1}^{\text{RBD}}.$

\section{Proposed DNN-based System}
\label{proposed_DNN_based_system}
In this study, an end-to-end DL-based precoding system is proposed in which the BS and users have fully connected individual networks (Fig. \ref{fig8}; red box). 
The receiver DNN abstracts the channel estimation and quantization process; the input is the received signals, $\mathbf{Y}_k^{\text{train}}\triangleq[\mathbf{y}_{1,k}^{\text{train}}, \cdots, \mathbf{y}_{L,k}^{\text{train}}]$, which correspond to the pilot transmission and the output is a quantization index (represented by binary vector). 
Subsequently, the output indices are fed back to the transmitter DNN to determine the precoding matrices. 
In Sections \ref{FC_DNN} and \ref{rx_tx_DNNs}, we first describe the overall structure and then the KD method is described in Section \ref{knowledge distillation}.

\subsection{Basic operation of a fully connected DNN}
\label{FC_DNN}

Denoting $D_n$ as the dimension of the $n$-th hidden layer of a fully connected network, the output $\mathbf{z}_n$ of the $n$-th hidden layer is defined as follows: 
\begin{align}\label{200720_0229_eq_1}
    &\mathbf{z}_n = a(\mathbf{W}_n\mathbf{z}_{n-1} + \mathbf{b}_n),
\end{align}
where $a(\cdot)$ is an element-wise activation function, $\mathbf{W}_n\in\mathbb{R}^{D_n\times D_{n-1}}$ is a weight matrix, and $\mathbf{b}_n\in\mathbb{R}^{D_n\times 1}$ is a bias vector.
In this study, we use a rectified linear unit (ReLU) activation, denoted by $a(x)=\max(0, x)$.
Denoting $t$ as the number of total hidden layers in a fully connected network, the output of DNN $\mathbf{z}_t$ is obtained by applying \eqref{200720_0229_eq_1} recursively; the activation function is not applied for the last recursion.
Let $\mathbf{\Theta}$ be the set of all the parameters of the corresponding fully connected network; the output can then be expressed as $\mathbf{z}_t = \text{FC}(\mathbf{z}_0; \mathbf{\Theta})$. 


\subsection{Proposed DL-based MU-MIMO system}
\label{rx_tx_DNNs}
For each user $k$, the received signal $\mathbf{Y}_k^{\text{train}}$ for channel estimation is first converted to a real vector, and then used as an input for the receiver DNN implemented at the user.
Let $\mathbf{r}_k^{\text{C}}$ be a complex vector obtained by stacking the columns of $\mathbf{Y}_k^{\text{train}}$ one after the other from the first to the last column.
Then, $\mathbf{r}_k^{\text{Re}} = [\text{Re}(\mathbf{r}_k^{\text{C}})^T \;\;\text{Im}(\mathbf{r}_k^{\text{C}})^T]^T$ is used as an input for the receiver DNN at user $k$.
Based on the operation described in Section \ref{FC_DNN}, a fully connected network $\text{FC}_k^{\text{Rx}}$ at user $k$ is used to estimate a real-valued output vector $\mathbf{u}_k\in\mathbb{R}^{B\times 1}$ as $\mathbf{u}_k=\text{FC}_k^{\text{Rx}}(\mathbf{r}_k^{\text{Re}}; \mathbf{\Theta}_k^{\text{Rx}})$. 
Each element of this vector is binarized using the sign function after an application of the hyperbolic tangent function to obtain the feedback information (Fig. \ref{fig8}; green box).
Denoting the $i$-th element of $\mathbf{u}_k$ as $[\mathbf{u}_k]_i$, the input-output relation for the receiver DNN of user $k$ can be represented using a mapping function $f_k^{\text{Rx}}(\cdot)$:
\begin{align}
    \label{200718_1423_eq_22}
    \hat{\mathbf{q}}_k &= f_k^{\text{Rx}}(\mathbf{Y}_k^{\text{train}}, \mathbf{\Theta}_k^{\text{Rx}}) = \text{sign}(\tanh(\text{FC}_k^{\text{Rx}}(\mathbf{r}_k^{\text{Re}}, \mathbf{\Theta}_k^{\text{Rx}})))
    \nonumber \\
    &= [\text{sign}(\tanh([\mathbf{u}_k]_1)), \cdots, \text{sign}(\tanh([\mathbf{u}_k]_B))].
\end{align}
where $\mathbf{\Theta}_k^{\text{Rx}}$ is the parameter set of a fully connected network included in the receiver DNN of user $k$ and $\hat{\mathbf{q}}_k$ is a vector of size $B$ with each element of $\hat{\mathbf{q}}$ being $1$ or $-1$.
Therefore, it has a one-to-one correspondence with the feedback index $q_k$ of the baseline limited feedback system. 


The transmitter DNN determines the concatenated precoding matrix  $\mathbf{V}=[\mathbf{V}_1 \cdots \mathbf{V}_K]$ by combining the feedback information from the users.
Because $\mathbf{V}$ is a complex matrix of size $M\times NK$, the output of the fully connected network $\text{FC}^{\text{Tx}}$ inside the transmitter DNN must have $2MNK$ real values.
They must be rearranged to formulate a complex matrix of size ${M\times NK}$ (any one-to-one corresponding mapping can be used for this arrangement), and subsequently each column of the matrix must be normalized.
Let $h$ be the corresponding function from the output of $\text{FC}^{\text{Tx}}$ to the rearranged complex matrix with a normalized column.
Then, the overall input-output relation of the transmitter DNN $f^{\text{Tx}}$ can be described as
\begin{align}
    \label{200720_1229_eq_1}
    \mathbf{V}=[\mathbf{V}_1 \cdots \mathbf{V}_K] 
    &= f^{\text{Tx}}(\hat{\mathbf{q}}_1, \cdots, \hat{\mathbf{q}}_K, P; \mathbf{\Theta}^{\text{Tx}})
    \nonumber \\
    &= h\left(\text{FC}^{\text{Tx}}(\hat{\mathbf{q}}_1, \cdots, \hat{\mathbf{q}}_K, P; \mathbf{\Theta}^{\text{Tx}})\right), 
\end{align}
where $\mathbf{\Theta}^{\text{Tx}}$ is the parameter set of a fully connected network included in the transmitter DNN.

Because our overall training problem aims to maximize the sum-rate of (\ref{eq5}), the loss function for training is defined as the minus sum rate: 
\begin{align}\label{200720_0300_eq_1}
    &\min_{\mathbf{\Theta}^{\text{Tx}}, \mathbf{\Theta}_1^{\text{Rx}}, \cdots, \mathbf{\Theta}_K^{\text{Rx}}} \quad
    L_{main}(\{\mathbf{\Theta}_k^{\text{Rx}}\}_{k=1}^{K},\mathbf{\Theta}^{\text{Tx}}),
\end{align}
\begin{align}\label{200807_1744_eq_4}
    &L_{main}(\{\mathbf{\Theta}_k^{\text{Rx}}\}_{k=1}^{K},\mathbf{\Theta}^{\text{Tx}})
    \nonumber \\
    &\quad = -\sum_{k=1}^{K}R_{k}\left(f^{\text{Tx}}(\{f_k^{\text{Rx}}(\mathbf{Y}_k^{\text{train}}; \mathbf{\Theta}_k^{\text{Rx}})\}_{k=1}^K, P; \mathbf{\Theta}^{\text{Tx}})\right).
\end{align}

To map the real-valued output vector $\mathbf{u}_k$ of each receiver DNN to a binary sequence of $1$ or $-1$, we can directly apply one of the common binarization operators (e.g., sign$(\cdot)$, etc.), as in (\ref{200718_1423_eq_22}). 
However, when these non-differentiable operators are used in an end-to-end learning process, the parameters of the receiver DNNs are not updated during the backward pass because of the vanishing gradient problem. 
To alleviate this problem, most studies employed the STE \cite{guo2020convolutional, liu2020efficient, Sohrabi2020, guo2020deep, Jang2020, guo2020dl, lu2019bit, bengio2013}, which replaces the binarization operator with a smooth differentiable function in the backward pass.
In other words, a smooth differentiable function layer is added in front of the binarization layer and the binarization layer is only exploited in the forward pass.
In this study, we exploit the hyperbolic tangent function as the smooth differentiable function, like in \cite{Jang2020} (Fig. \ref{fig8}).
Therefore, for the binarization layer, we exploit the approximated gradient of $\text{sign}(\tanh(z))$ in the backward pass as follows:
\begin{align}
    \label{200720_0713_eq_1}
    \nabla_{\Theta_k^R}\text{sign}(\tanh(z))\approx\nabla_{\Theta_k^R} \tanh(z).
\end{align}
By using an STE, we can train the receiver DNNs and transmitter DNN in an end-to-end manner.

However, according to \cite{zhuang2018, zhuang2019, zhuang2020}, the approximated gradient (\ref{200720_0713_eq_1}) of the binarization function adds a noise when updating the DNN parameters because of an incorrect update direction.
Therefore, the receiver DNNs can be trapped at a poor local minima, and consequently, the performance may be degraded.
Moreover, this error can be propagated to the transmitter DNN because of the end-to-end learning.
The best solution to overcome the noisy gradient problems is to provide ``lossless gradients'' to receiver DNNs. 
To achieve this, we propose a novel joint training method using KD.

\subsection{Joint training method with KD}
\label{knowledge distillation}
KD \cite{hinton2014} 
is a method that allows a deep teacher network to distill noisy knowledge and transfer the distilled knowledge to a shallow student network to improve the training performance of the student network. 
In our case, if we connect the deep teacher network (we denote it as the auxiliary transmitter DNN) to the outputs of the layer immediately before the tanh layer of the receiver DNNs (blue dotted line in Fig. \ref{fig8}), we can directly provide lossless gradients to the receiver DNNs during the end-to-end learning because the auxiliary DNN ignores the binarization layer (Fig. \ref{fig8}; blue box).
The input-output relation of the auxiliary transmitter DNN is defined as follows:
\begin{align}
    \label{200807_2051_eq_1}
    \mathbf{V}
    &= f^{\text{Tx}}(\mathbf{u}_1, \cdots, \mathbf{u}_K, P; \mathbf{\Theta}_{aux}^{\text{Tx}})
    \nonumber \\
    &= h\left(\text{FC}^{\text{Tx}}(\mathbf{u}_1, \cdots, \mathbf{u}_K, P; \mathbf{\Theta}_{aux}^{\text{Tx}})\right),
\end{align}
where the auxiliary DNN has the same network structure as the original transmitter DNN described in \eqref{200720_1229_eq_1}. The primary difference is that the auxiliary DNN takes the output vectors of the receiver DNNs before binarization, i.e., $\mathbf{u}_k=\text{FC}_k^{\text{Rx}}(\mathbf{r}_k^{\text{Re}}, \mathbf{\Theta}_k^{\text{Rx}})$, for $k=1\cdots, K$, as the input values.

The structure composed of the receiver DNNs and auxiliary transmitter DNN can also be considered as a single fully connected network (the overall network is connected with blue dashed lines in Fig. \ref{fig8}). 
Therefore, similar to \eqref{200720_0300_eq_1}-\eqref{200807_1744_eq_4}, the optimization problem with the auxiliary DNN is defined as follows: 
\begin{align}\label{200807_2133_eq_3}
    &\min_{\mathbf{\Theta}_{aux}^{\text{Tx}}, \mathbf{\Theta}_1^{\text{Rx}}, \cdots, \mathbf{\Theta}_K^{\text{Rx}}} \quad L_{aux}(\{\mathbf{\Theta}_k^{\text{Rx}}\}_{k=1}^{K},\mathbf{\Theta}_{aux}^{\text{Tx}}),
\end{align}
\begin{align}\label{200807_1744_eq_5}
&\!\!\!\!L_{aux}(\{\mathbf{\Theta}_k^{\text{Rx}}\}_{k=1}^{K},\mathbf{\Theta}_{aux}^{\text{Tx}})
\nonumber \\
&\!\!\!\!\!\!\!\!\quad= -\sum_{k=1}^{K}R_{k}\left(f^{\text{Tx}}(\{\text{FC}_k^{\text{Rx}}(\mathbf{r}_k^{\text{Re}}; \mathbf{\Theta}_k^{\text{Rx}})\}_{k=1}^K, P; \mathbf{\Theta}_{aux}^{\text{Tx}})\right).
\end{align}

\begin{algorithm}[t]
\caption{$\textrm{Joint training with auxiliary DNN}$}
\begin{algorithmic}[1]
\State \textbf{Initialize} $\mathbf{\Theta}_k^{\text{Rx}}$, $\mathbf{\Theta}_{aux}^{\text{Tx}}$, $\mathbf{\Theta}^{\text{Tx}}$;
\For{$l=1:num\_iterations$}
	\State \textbf{Generate} channel data of each user as much as mini-batch ${\mathcal{B}}$;
	\State \textbf{Update} $\text{FC}_k^{\text{Rx}}(\mathbf{\Theta}_k^{\text{Rx}})$, $f^{\text{Tx}}(\mathbf{\Theta}_{aux}^{\text{Tx}})$ by minimizing  $L_{aux}$;
	\State \textbf{Update} $f_k^{\text{Rx}}(\mathbf{\Theta}_k^{\text{Rx}})$, $f^{\text{Tx}}(\mathbf{\Theta}^{\text{Tx}})$ by minimizing  $L_{main}$;
\EndFor
\end{algorithmic}
\label{algo1}
\end{algorithm}

%
%

To utilize the auxiliary transmitter DNN in the training phase, we employ a joint training method that alternatively trains the original and auxiliary transmitter DNN.
The detailed joint training procedure is described in Algorithm \ref{algo1}. The original and auxiliary DNN are sequentially updated for each iteration; note that an STE is used in step 5, but is not used in step 4.
Consequently, using lossless gradients, the auxiliary DNN guides the original transmitter and receiver DNNs with useful information for generalization, and the corresponding joint training prevents them from being trapped at a poor local minimum which occurs when the network is trained from scratch \cite{zhuang2018, zhuang2019}.

In a general DL framework for the KD, the key issue is to effectively transfer the useful knowledge to a shallow student network using a deep teacher network. 
However, most layers of the shallow student network restrict the knowledge transfer  
because they generally have different structures from those of the deep teacher network.
Fortunately, our study has only one bottleneck (binarization layer) at the end of the receiver DNNs. In other words, the shallow student network (the original transmitter DNN) and deep teacher network (the auxiliary transmitter DNN) have identical structures, except for the tanh function and the binarization layer (both tanh and binarization layers are not used in the deep teacher network as they cause the vanishing gradient problem). 
Therefore, our case is suitable for the use of KD. 
To the authors' knowledge, the concept of KD has not been considered in previous studies of DL-based quantization and precoding using STEs, although the STEs can cause a noisy gradient problem when emulating the conventional codebook-based quantization.

%
%

Lastly, we compare the computational overhead between the proposed DNN-based approach and the conventional codebook-based linear precoding in terms of the floating-point operations (FLOPs). 
Because the trained transmitter DNN leads to a one-to-one mapping from the feedback information to the corresponding precoding matrices, a lookup table implementation is possible for the transmitter, as discussed in \cite{Jang2020}. Therefore, the overall computational overhead depends only on the receiver DNNs. 
The total number of FLOPs in the proposed DNN-based approach is  $10MN(16LN+360MN-9)+B(40MN-1)$.
    The FLOPs of the baseline scheme can be calculated by counting the real multiplications and additions in the basic matrix algebra: 1) multiplication of $m\times n$ and $n\times p$ real matrices requires $mnp$ multiplications and $m(n-1)p$ additions, 2) inversion and determinant operations of an $n\times n$ matrix are approximated by the multiplication of two $n\times n$ matrices for simplicity. Then, the number of FLOPs for channel estimation and quantization in the baseline scheme is approximated as  $2^B(32M^2N+16MN^2+16N^3-4M^2-4MN-8N^2)+16M(LM+LN+MN)+16L^2(M+N+L)-4L(M+N)-8L^2-8MN$. 
    The increasing rate of the number of FLOPs with respect to $L$ is larger for the proposed scheme than for the baseline schemes, whereas the increasing rate with respect to $B$ is larger for the baseline schemes.  
    For moderate values of $L$, $M$, $N$, and $B$, the FLOPs of the proposed and baseline schemes are comparable. 
    For example, if $L=8$, $M=8$, $N=2$, and $B=6$, then $\text{FLOPs}^{\text{baseline}}\approx 310,336$ and $\text{FLOPs}^{\text{Proposed}}=471,040$.

\section{Numerical Results and Conclusions}
\label{NR}
\subsection{Simulation setup}
The total throughput of the system is evaluated using the sum rate, $\sum_{k=1}^{K}R_{k}$.
For the baseline BD \cite{Ravindran2008} and RBD \cite{Stankovic2008} schemes, the expected signal-to-interference-plus-noise ratio (SINR) proposed in \cite{Min2016} is adopted as the distance measure in \eqref{eq2}.
To obtain an appropriate codebook, the Lloyd algorithm \cite{Xia2006} is used for each simulation setup.
For the pilot, we assume $P_{\text{train}}=10$ and $L=M$ for all the simulations.
The optimal value of $\alpha_{\text{BD}}$ is numerically determined based on a brute-force search for each SNR for each simulation setup.

The end-to-end supervised learning is performed using the following setup. 
Each receiver DNN consists of a three-layer fully connected network with dimensions of $40MN$, $30MN$, and $20MN$.
The transmitter DNN consists of a three-layer fully connected network with dimensions of $20MNK$, $30MNK$, and $40MNK$.
The auxiliary transmitter DNN has the same structure (but different parameters $\mathbf{\Theta}_{aux}^{\text{Tx}}$) as the transmitter DNN.
We set a batch size of $1000$ with the Adam optimizer \cite{kingma2014}.
The training was conducted for $50000$ iterations with an initial learning rate of $2\times10^{-4}$, reducing the rate by $0.1$ times, at the $30000$th and $40000$th iterations.
We randomly generate a training dataset for each training iteration. 
For testing the trained DNNs, we evaluate $100,000$ fixed test datasets that are independently sampled. 
In addition, we use $100,000$ fixed validation datasets that are independently sampled to determine the best performing model.
The algorithms were implemented using TensorFlow 1.8.0.
Simulation codes are available at https://github.com/kyeongbokong/Knowledge-Distillation-aided-End-to-End-Learning-for-Linear-Precoding-in-Multiuser-MIMO-Downlink-Sys.
Note that we have to train the DNNs separately for different values of $K$.

\begin{figure}
    \centering
    {\epsfig{file=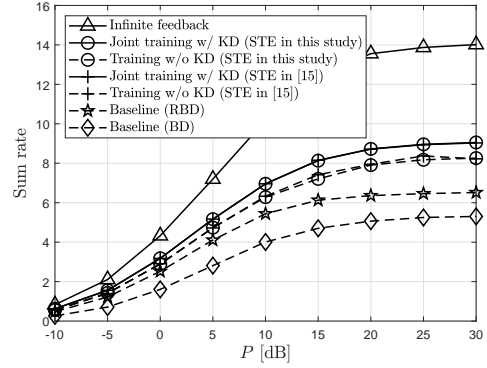, width=7.0cm}}
    \caption{Sum rate vs. $P$ (dB). $M=8$, $N=2$, $K=4$, and $B=6$}
    \label{fig6} 
\end{figure}

\begin{figure}
    \centering
    {\epsfig{file=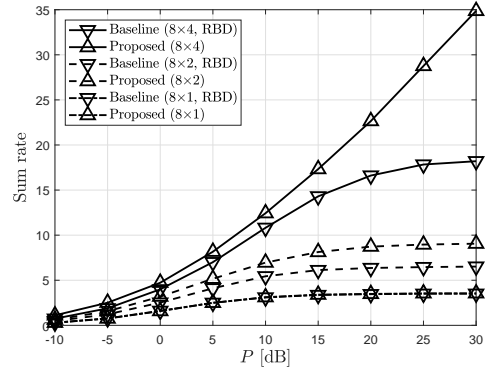,width=7.0cm}}
    \caption{Sum rate vs. $P$ (dB). The total number of feedback bits is fixed, i.e., $BK=24$ where $K=M/N$ }
    \label{fig2}
\end{figure}

\subsection{Simulation results}
Fig. \ref{fig6} shows the performance enhancement achieved by the proposed DNN-based precoding.
The sum rates of the proposed joint training with KD are generally higher than those of the training without KD; training without KD corresponds to the extension of \cite{Jang2020} as depicted in Fig. \ref{fig8}.
Both DNN-based precodings generally have higher sum rates than the conventional limited-feedback-based BD and RBD.
Fig. \ref{fig6} also depicts the sum rates achieved using a sigmoid-adjusted STE considered in \cite{Sohrabi2020} (dashed line with $+$ markers). 
The work in \cite{Sohrabi2020} did not consider the KD. 
Therefore, we applied our joint training method to the sigmoid-adjusted STE in \cite{Sohrabi2020} for KD to improve the training performance (solid line with $+$ markers). 
The corresponding results demonstrate that our joint training with KD can be successfully applied to different STEs in the literature as long as it uses a quantization or binarization layer. 
The solid line with upper triangle markers shows the sum rate of RBD achieved with an infinite feedback rate; i.e., the sum rate achieved when the transmitter can exactly know the estimated channel $\bar{\mathbf{H}}_{k}$. 
The corresponding result provides an upper bound sum rate for a limited-feedback-based RBD and shows the effectiveness of the proposed DL-based precoding.

\begin{figure}
    \centering
    {\epsfig{file=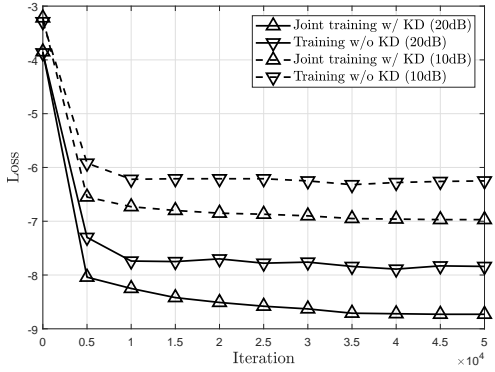,width=6.5cm}}
    \caption{Loss vs. Iteration. $M=8$, $N=2$, $K=4$, and $B=6$}
    \label{Training_figure}
\end{figure}

In Fig. \ref{fig2}, the sum rates are compared by increasing the number of receive antennas where $M$ is fixed at $8$.
The rate gap between the proposed and baseline schemes increases with the number of receive antennas at moderate and high $P$.
The relatively poor performance of the RBD when $N>1$ is due to difficulties in optimizing the limited feedback system.
Specifically, it is difficult to obtain a distance measure and quantization codebook that jointly optimizes the quantization and precoding when $N>1$. 
Moreover, the columns of each precoding matrix of the baseline BD and RBD are represented by the orthogonal vectors, because it is a good choice with the perfect CSIT.
However, with limited feedback, an optimal precoding matrix may not have orthogonal columns.
In fact, our DL-based solution for precoding matrices does not have orthogonal columns which indirectly indicates the inefficiency in selection of orthogonal vectors with limited feedback.
This enables our proposed scheme to achieve a higher sum rate gap when a large number of independent layers is allocated to each user.
Lastly, to verify stable convergence for each training, we have performed ten independent trials for a joint training with KD. 
    For example, at 30 dB, the loss converges to a sample solution of $-9.036$ with errors of $\pm0.003$ for all independent trials. 
    In Fig. \ref{Training_figure}, we also demonstrate that the loss decreases uniformly and converges rapidly as the training progresses.

%

%


\end{document}